\documentclass[aps,prl,twocolumn,amsmath,amssymb,groupedaddress,showpacs]{revtex4-1}
\usepackage{graphicx}
\usepackage{dcolumn}
\usepackage{revsymb}

\newcommand*{\term}[3]{\mbox{$^{#1}{\rm #2}_{#3}$}}
\newcommand{\bra}[1]{\left\langle #1\right|}
\newcommand{\ket}[1]{\left|#1\right\rangle}

\begin{document}


\title{Controlled Production of Sub-Radiant States of a Diatomic Molecule in an Optical Lattice}


\author{Yosuke Takasu$^1$}
\author{Yutaka Saito$^1$}
\author{Yoshiro Takahashi$^{1,2}$}
\author{Mateusz \surname{Borkowski}$^{3}$} 
\author{Roman \surname{Ciury{\l}o}$^{3}$}
\author{Paul S. Julienne$^{4}$}
\affiliation{$^1$Department of Physics, Graduate School of Science, Kyoto University, Kyoto 606-8502, Japan\\
$^2$CREST, Japan Science and Technology Agency, Chiyoda-ku, Tokyo 102-0075, Japan\\
$^3$Instytut Fizyki, Uniwersytet Miko\l{}aja Kopernika, 
  ul. Grudzi\c{a}dzka 5/7, 87--100 Toru\'n, Poland.\\
$^4$Joint Quantum Institute, National Institute of Standards and Technology,
and the University of Maryland,
100 Bureau Drive, Stop 8423, Gaithersburg, Maryland 20899-8423, USA
}


\date{\today}

\begin{abstract}
We report successful production of sub-radiant states of a two-atom system in a three-dimensional optical lattice starting from doubly occupied sites in a Mott insulator phase of a quantum gas of atomic ytterbium.
We can selectively produce either sub-radiant $1_g$ state or super-radiant $0_u$ state by choosing the excitation laser frequency.
The inherent weak excitation rate for the sub-radiant $1_g$ state is overcome by the increased atomic density due to the tight-confinement in a three-dimensional optical lattice.
Our experimental measurements of binding energies, linewidth, and Zeeman shift confirm observation of sub-radiant levels of the $1_g$ state of the $\text{Yb}_2$ molecule.
\end{abstract}


\maketitle
Spontaneous emission of dipole radiation from two identical atoms is a fundamental process theoretically studied by R. H. Dicke, who introduced the important concepts of super-radiance and also sub-radiance~\cite{Dicke}.  
While both of these correspond to quantum superpositions of product states with a ground state in one atom and an excited state in another, the spontaneous emission is enhanced (suppressed) in the super-radiant (sub-radiant) state due to constructive (destructive) interference.
Although the phenomenon of super-radiance is widely observed in various physical systems, sub-radiance is not well studied~\cite{[{Slightly reduced spontaneous decay rate was reported in two-ions system. }][]Subradiant_Ion}.
The difficulty of creating and studying the sub-radiant state naturally comes from its reduced radiative interaction.

In this Letter, we report our successful controlled production of a sub-radiant state of a two-atom system in single quantum states of both internal ro-vibronic and external center-of-mass degrees of freedom, starting from doubly occupied sites in a Mott insulator phase of an atomic quantum gas in a three-dimensional optical lattice.
For the case we study, the sub-radiant and super-radiant states correspond respectively to $1_g$ and $0_u$ states, labeled by the standard molecular Hund's case (c) notation.
By using a two-electron atom with no structure in the ground state, we realize an ideal situation for observing and studying a sub-radiant $1_g$ state.
In addition, the inherent weak excitation rate is overcome by tight-confinement of two atoms in a three-dimensional optical lattice.
A suppressed spontaneous emission rate is confirmed by the linewidth measurement as one of the important evidences of the sub-radiance.
We can selectively produce either a sub-radiant $1_g$ or super-radiant $0_u$ state by choosing the excitation laser frequency.
In addition, observed binding energies are in excellent agreement with theoretical analysis.
While it might be possible to populate these sub-radiant molecular states by some simpler scheme in a uncontrolled or non-selective way, it should be difficult to study the system with high precision and controllability.
We believe that our highly controlled production scheme will open the door to a number of interesting experiments.
For example, the inherent sub-kHz spectrum linewidth of the optical transition from the ground state to these sub-radiant molecular states is straightforwardly applicable to a high-resolution photoassociation spectroscopic study of a quantum many body state in an optical lattice.
In addition, our scheme will offer a novel possibility of creating a long-lived electronically excited molecular quantum gas in which the collisional stability can be studied.

The studied sub-radiant state of two-atom system corresponds to a diatomic molecular state with an inversion symmetry, which is one of the fundamental symmetries exhibited by a homonuclear diatomic molecule. 
Quantum states whose wavefunction retain their sign upon inversion of electron coordinates are labelled \emph{gerade}, as opposed to \emph{ungerade} states, whose wavefunction
changes sign after inversion.
We write two degenerate molecular excited states:
$\ket{A(\term{}{P}{})B(\term{}{S}{})}$ and
$\ket{A(\term{}{S}{})B(\term{}{P}{})}$,
where S and P represent the ground and excited states, respectively and A and B represents two atoms. 
As molecular eigenstates we consider the symmetrized molecular states 
\begin{equation}
  \ket{\Psi_\pm} = \frac{1}{\sqrt{2}} \left( \ket{A(\term{}{P}{})B(\term{}{S}{})}
    \pm \ket{A(\term{}{S}{})B(\term{}{P}{})} \right),
\end{equation}
where, the $\ket{\Psi_{+}}$ state is of \emph{ungerade} symmetry and the $\ket{\Psi_{-}}$ state is \emph{gerade} because the atomic excited state considered here itself is \emph{ungerade}.
The fact that the dipole operator, whose matrix elements determine the transition width, is itself \emph{ungerade} leads to the Laporte selection rule~\cite{Herzberg} which states that optical dipole transitions are only possible between states of opposite inversion symmetry.  
The \emph{gerade}-\emph{gerade} (\emph{g-g}) transition is, therefore, forbidden, whereas a \emph{gerade}-\emph{ungerade} (\emph{g-u}) one is allowed, having a natural linewidth twice the bare atomic one $\Gamma_{\rm at}$.
These are nothing but the diatomic molecular realizations of phenomena of sub-radiance and super-radiance.

It is noted that a retardation effect caused by the finite speed of interaction propagation, although weak, can lead to the breakdown of the Laporte rule.
An example from classic spectroscopy is the well-known Lyman-Birge-Hopfield band of the nitrogen molecule~\cite{N2-1,N2-2}, which is the first confirmed molecular \emph{gerade}-\emph{gerade} transition~\cite{N2-AJ}.
Consider a diatomic molecule immersed in an electromagnetic field of a laser beam. 
If the two atoms are confined closely the electric field `felt' by the two is approximately the same.
In this case, we expect considerable suppression of the spontaneous emission or optical excitation.
In other words, it is very difficult to create a sub-radiant state by optical excitation from the ground state.
In reality, however, the two atoms in molecular states close to the dissociation are weakly coupled and relatively far apart. 
Therefore, there is a phase difference in the electric fields experienced by them.
When calculating the transition rate to the excited state from the ground state of {$\ket{A(S)B(S)}$}, we have to include this phase difference~\cite{Power1967} and replace the usual transition matrix element by
\begin{equation}
  \left|{\bra{A(S)B(S)}} \hat d_{A} + \hat d_{B} \exp(-i \vec k \cdot \vec R) \ket{\Psi_\pm} \right|^2 \, ,
\end{equation}
where $\vec R$ is the distance between the atoms and $\vec k$ is the incident photon's wavevector.
When we consider the spontaneous emission, this has to be averaged over all possible photonic states. 
The transition widths {$\gamma$} depend on the projection $\Lambda$ of the orbital angular momentum on the internuclear axis ($\Sigma$ states have {$\Lambda = 0$} while $\Pi$ states have {$|\Lambda| = 1$}) and are~\cite{Stephen1964, Hutchinson, Meath1968, Power1967, Machholm.AE1g}
\begin{eqnarray}
  \gamma(\Sigma) & = & \Gamma_{\rm at} \left(1 \pm \frac{3}{v^3}[-v \cos v + \sin v] \right) \label{eq:gsigma}\\
  \gamma(\Pi)& = & \Gamma_{\rm at} \bigg(1 \pm \frac{3}{2v^3} [v \cos v - (1-v^2)\sin v] \bigg) \label{eq:gpi}
\end{eqnarray}
where {$v = kR$} and ``+'' is for allowed \emph{g-u} transitions, while ``-'' applies to the forbidden \emph{g-g} and \emph{u-u} transitions.
These expressions for $\Lambda=0$ and $|\Lambda|=1$ states translate directly to those for $\Omega = 0$ and $|\Omega| = 1$ states in Hund's case (c)~\cite{Mies78, Ciurylo04} which is relevant to the molecular states in this work, where $\Omega$ is the projection of the total electronic angular momentum  $j$ on the molecular axis.
Note that in the limit of small internuclear distances ($v \leadsto 0$) these expressions lead to $2\Gamma_{\rm at}$ for the allowed \emph{g-u} transitions and zero for the forbidden \emph{g-g} and \emph{u-u} transitions.
At infinite separations, $v \leadsto \infty$, these widths become $\Gamma_{at}$, exactly as we would expect for uncorrelated atoms.

Although Ref.~\cite{Lett,Stwalley} studies deeply bound levels of alkali dimer, the alkali species are complicated by \emph{g-u} mixing in the ground state induced by nuclear spin.
Having a non-degenerate ground state is an important prerequisite for the clear demonstration of sub-radiance. This is satisfied in our work by using a two-electron atom of ytterbium (${}^{174}$Yb) where the ground state is ${}^1\text{S}_0$. 

\begin{figure}
\includegraphics[width=0.45\textwidth]{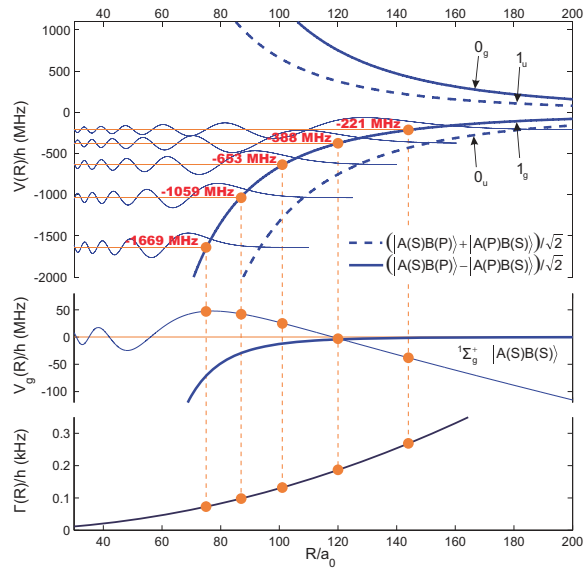}

  \caption{(color online) Bound states in the excited \term{}{1}{g} state in $J=1$. 
    \emph{Top panel}: energies, outer turning points and wavefunctions
    of the bound states in the investigated range. \emph{Middle panel}: 
    ground state wavefunction at a collision energy of $\varepsilon_r/k_B = $~1~$\mu$K.
    Note that the outer turning point of the -388~MHz state is almost exactly
    on the node of the ground state wavefunction, making it unobservable.
    \emph{Bottom panel}: natural linewidth $\Gamma$(\term{}{1}{g})
    of the \emph{g-g} transition 
    expected in retardation theory (see Eq.~(\ref{eq:gpi})). 
    Note that for most transitions this width is less than 200~Hz.
    \label{fig:states}}
\end{figure}

Since the molecular levels we study have binding energies much smaller than the fine structure splitting in the excited states, they are well described by Hund's coupling case (c).
The excited molecular states that are asymptotically connected to the {${}^{1}\text{S}_0 + {}^{3}\text{P}_1$} separated atom limit are the doubly degenerate $1_g$ and $1_u$ ({$|\Omega|=1$}) states and the $0_g$ and $0_u$ ({$\Omega=0$}) nondegenerate states (see Fig.~\ref{fig:states}).
The subscript $g$ ($u$) denotes \emph{gerade} (\emph{ungerade}) symmetry.
The $0_u$ and $1_g$ states have long-range attractive molecular potentials that support respective super-radiant and sub-radiant bound states.
We present the interaction potential $V^{1_g}$ of two atoms in the $1_g$ sub-radiant state as~\cite{Mies78, Ciurylo04, Borkowski.556PAtheory}
\begin{align}
V(R)=&-\frac{C^{1_g}_3 t(R)}{R^3}-\left(1-\frac{\sigma_{1_g}^6}{R^6}\right)\frac{C_6^{1_g}}{R^6}-\frac{C_8^{1_g}}{R^8} \notag\\
     &+\frac{\hbar^2\left\{J(J+1)\right\}}{ 2\mu R^2}  \label{potential}
\end{align}
where $\hbar$ is the Planck constant divided by $2\pi$, $\mu$ is the reduced mass of the Yb atom pairs, $t(R)=\cos v+v \sin v-v^2\cos v$, and $J$ is the quantum number for total angular momentum.
The resonant dipole-dipole interaction coefficient $C^{1_g}_3$ is most conveniently described in terms of the atomic transition linewidth: $C^{1_g}_3 = (3/4)\Gamma_{at}(\lambda/2\pi)^3$ where $\lambda$ is the wavelength of the ${}^{1}\text{S}_0-{}^{3}\text{P}_1$ transition. 
From a simple theoretical argument, the value of $C^{1_g}_3$ is one half of $C^{0_u}_3$ which is the resonant dipole-dipole interaction coefficient for the $0_u$ state. 
The terms including $C^{1_g}_6$ and $C^{1_g}_8$ represent the van der Waals interaction.
The {$\sigma_{1_g}^6$} parameter characterizes the strength of the short-range interaction~\cite{Borkowski.556PAtheory}.

The experimental procedure is as follows.
First, we generate a ${}^{174}\text{Yb}$ Bose-Einstein Condensate (BEC)~\cite{Takasu.YbBEC}.
Then we adiabatically ramp up a three-dimensional optical lattice potential using three mutually orthogonal laser beams with the wavelength of $532~\text{nm}$~\cite{Fukuhara.Mott}.
The typical potential depth of the optical lattice is about $22E_R$, in which $E_R=0.2~\mu \text{K}$ is the recoil energy for the Yb atom.
At this lattice depth, the system forms a Mott insulator where the atoms are well localized at each lattice site up to the filling of three atoms per site, depending on the total atom number.
Then, the linearly polarized photoassociation light which is red-detuned from the ${}^{1}\text{S}_0 - {}^{3}\text{P}_{1}$ transition of Yb atoms is focused on the atoms in the optical lattice with a beam waist of $70~\mu\text{m}$.
The typical pulse has a duration of $100~\text{ms}$ and a power of $200~\mu\text{W}$.
The photoassociation light induces trap loss of the atoms when it is resonant with a molecular bound state in the excited state.
The number of the atoms remained in the trap is measured with an absorption imaging technique.
The probe light has a duration of $\sim 100~\mu\text{s}$ and a power of $\sim 100~\mu\text{W}$.
The residual magnetic field is reduced to less than $0.5$ $\text{G}$.

\begin{figure*}
\begin{center}
\includegraphics[width=150mm]{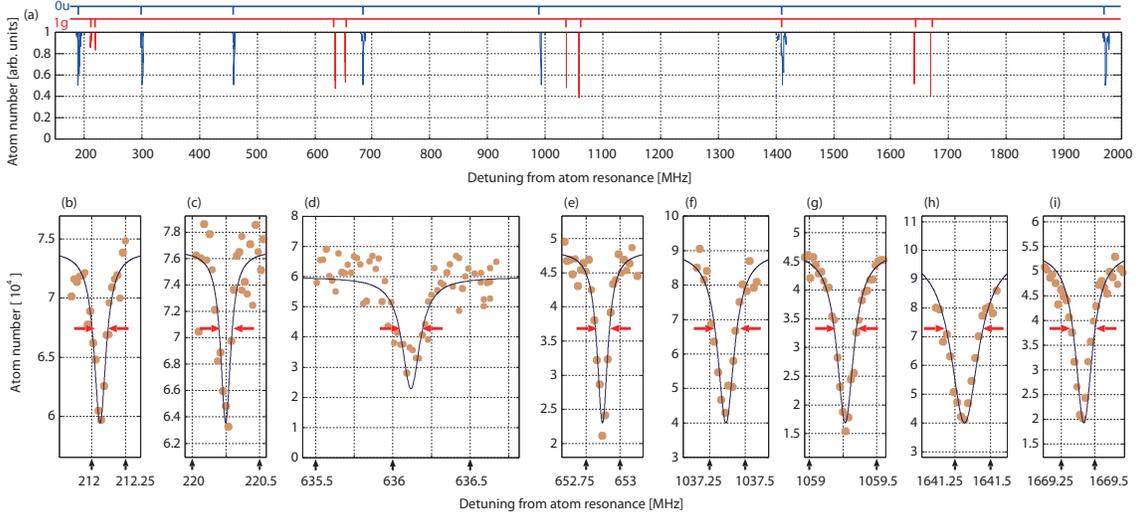}
\end{center}
\caption{\label{fig:spectrum}(color online) (a): All photoassociation spectra of the $1_g$ states we observed. The $0_u$ states are also shown and are taken from Ref.~\cite{Tojo.556PA}. The horizontal and vertical axes are the frequency difference of excitation laser from the atom resonance ${}^{1}\text{S}_0$ - $^{3}\text{P}_1$ and the number of remaining atoms, respectively. (b)-(i): Spectra of the $1_g$ states. The solid line is a fit of the Lorentz function.
Arrows show full widths at half maximum of the Lorentz function and the widths are (b) 0.108(17) MHz, (c) 0.088(24) MHz, (d) 0.141(25) MHz, (e) 0.095(8) MHz, (f) 0.142(22) MHz, (g) 0.150(8) MHz, (h) 0.236(41) MHz, (i) 0.151(15) MHz, respectively}
\end{figure*}%
With this procedure, we found eight new resonances within the range of 2 GHz red-detuning, in addition to the already observed photoassociation resonances associated with the $0_u$ state.
Figure~\ref{fig:spectrum} shows spectra of the newly found photoassociation resonances.
The new resonances are quite different from those of $0_u$ and are successfully assigned as levels of the $1_g$ states (see Table~\ref{table:0u1g}), as explained in detail below.
The measured binding energies in the Table I are compensated for the light shifts due to the photoassociation light by measuring the intensity dependence of the binding energies and extrapolating to zero intensity.
\begin{table}[t]
\caption{\label{table:0u1g}Measured and calculated binding energies ($E_b$) for sub-radiant molecular states, where $v$ and $J$ are the vibrational and rotational quantum numbers of the states.
$v'=v'_D-v$ is numbered from the dissociation limit. Values of calculated Condon points ($R_C$) are also shown, where $a_0$ is the Bohr radius.}
\begin{ruledtabular}
\begin{tabular}{c|c|cc|c|cc}
 & \multicolumn{3}{c|}{$J=1$} & \multicolumn{3}{c}{$J=2$} \\
\hline
 & \multicolumn{1}{c|}{Exp.} & \multicolumn{2}{c|}{Theory} & \multicolumn{1}{c|}{Exp.} & \multicolumn{2}{c}{Theory}  \\
$v'$ & $E_b$ & $E_b$ & $R_C$ & $E_b$ & $E_b$ & $R_C$ \\
 & (MHz) & (MHz) & ($a_0$) & (MHz) & (MHz) & ($a_0$) \\
\hline
9 & -221.23(3) & -221.13 & 142 & -212.08(2) & -212.26 & 143 \\
10 & \multicolumn{1}{c|}{not found} & -388.38 & 118 & \multicolumn{1}{c|}{not found} & -376.05 & 119 \\
11 & -652.91(3) & -652.92 & 100 & -636.33(18) & -636.29 & 100 \\
12 & -1059.33(7) & -1059.36 & 85.3 & -1037.53(7) & -1037.48 & 85.6 \\
13 & -1669.45(8) & -1669.51 & 73.9 & -1641.41(16) & -1641.37 & 74.0 \\
\end{tabular}
\end{ruledtabular}
\end{table}%

It should be also noted that we could not observe the $1_g$ resonances in a weak harmonic trap.
When we confine atoms in the vibrational ground state of an optical lattice, however, the effective local density is increased by the ratio of one over the volume of the ground state of the optical lattice to the number of atoms over the whole trap volume of the free gas.
In addition, by using the quantum gas, namely, two atoms in the lowest vibrational state in an optical lattice, the observed spectra do not suffer from the broadening due to the $k_B T$ spread in the case of a free thermal gas with a temperature $T$, where $k_B$ is the Boltzmann constant.
These effects result in the conversion of a weak free-bound transition into a rather stronger bound-bound transition~\cite{MTJ, MJ}.

We carefully confirm that the new resonances originate from the sub-radiant $1_g$ state.
The first important evidence is the spectrum linewidth.
As shown in Fig.~\ref{fig:spectrum}(b)-(i) the widths of the new resonances are narrower than the theoretical value of $2\times 182~\text{kHz}$ of the super-radiant $0_u$ state~\cite{Tojo.556PA}, and especially the linewidths in Fig.~\ref{fig:spectrum}(c) and (e) are less than $100~\text{kHz}$, which are even narrower than the atomic linewidth of $182~\text{kHz}$.
From this feature we can exclude the possibility that the new resonance originates from single atoms or the super-radiant $0_u$ state.
Since the linewidth of our photoassociation laser is measured to be almost the same as the measured photoassociation spectrum width, the intrinsic width of the new resonance could be much narrower,  consistent with the theoretically expected value of $0.2~\text{kHz}$ for the sub-radiant $1_g$ state (see Fig.~\ref{fig:states}).

Next, we confirm that the observed binding energies well fit with the eigenvalues of the potentials in Eq.~(\ref{potential}) as shown in Table I.
They consist of two series of $J=1$ and $J=2$.
No resonances are assigned to $J=0$ and $J>2$.
The fitting parameters are $C_3^{1_g}$=$0.5\times 0.193706 \ E_h \ a_0^3$, $C_6^{1_g}$=$2283.6 \ E_h \ a_0^6$, $C_8^{1_g}$=$748788 \ E_h \ a_0^8$, $\sigma_{1_g}$=$8.6006 \ a_0^{6}$, where $a_0$ is Bohr radius and $E_h$ is Hartree energy.
The value of $C_3^{1_g}$, which is expected as one half of $C_3^{0_u}$, is in good agreement with $C_3^{0_u}$=$0.194886 \ E_h \ a_0^3$~\cite{Borkowski.556PAtheory}, which supports that the new resonances originate from the sub-radiant $1_g$ state.
In our experiment, no resonances with the Condon points around 120 $a_0$ are observed, indicated as "not found" in Table I, since the Condon point is quite close to the node of the ground state wavefunction and therefore the transition rate is suppressed~\cite{Jablonski1945, Boisseau}. 
The interatomic potential and the bound state energies of the measured states are shown in the upper panel of Fig.~\ref{fig:states}.
The Condon points for the transitions to these states range from 75~$a_0$ to 144~$a_0$. According to the theory, these correspond to natural linewidths $\Gamma$ ranging from 100 to 250 Hz and thus are extremely weak.

\begin{figure}
\includegraphics[width=0.45\textwidth]{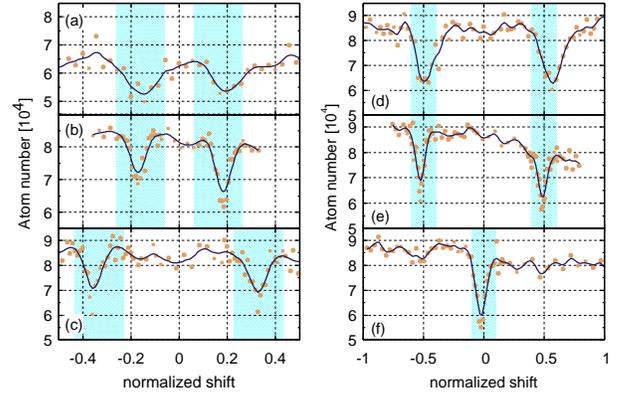}
\caption{\label{fig:Zeeman}
 (color online) Zeeman spectra of $v'=9$, $J=2$ resonance ((a), (b), (c)) and of $v'=9, J=1$ resonance((d), (e), (f)) under a magnetic field applied along the x- axis((a) and (d)), y-axis((b) and (e)), and z-axis((c) and (f)).
The frequency shifts shown here are normalized by the Zeeman shift of atoms {$g_{Ja}\mu_B B$}.
Shadow areas indicate the expected theoretical values of {$m_j/(J(J+1))$} (see Eq.~({\ref{eq:ZeemanShift}}) in the text), which are $1/6$ for (a),(b), $1/3$ for (c), $1/2$ for (d), (e) and 0 for (f), respectively. 
}
\end{figure}

Third, we investigate the Zeeman shift of the new resonance.
In Hund's case (c), the Zeeman splitting energy $E_{\text{Z}}$ by an external magnetic field $B$ is calculated within first-order perturbation theory~\cite{Brown} as 
\begin{equation}
E_{\text{Z}}=g_{mol} \mu_B m_J B=\frac{\Omega^2}{J(J+1)}g_{Ja} \mu_B m_J B \label{eq:ZeemanShift},
\end{equation}
where  $g_{Ja} \sim 1.5$ is Lande's g factor of Yb atoms in the ${}^{3}\text{P}_1$ states, and $\mu_B$ is the Bohr magnetic moment.
Here $|\Omega|=1$ for the $1_g$ state, and therefore the $1_{g}$ states have $g_{mol}=g_{Ja}/2$ for $J=1$ and $g_{mol}=g_{Ja}/6$ for $J=2$.
We experimentally measure the Zeeman shifts under an external magnetic field.
The spectra around -212.08 MHz and -221.229 MHz under different magnetic field directions are shown in Fig.~\ref{fig:Zeeman}.
We measure three kinds of spectra in order to study the Zeeman shifts as well as the selection rules of the transition.
An external magnetic field is applied along the x, y, or z axis, where x is the axis of polarization of photoassociation light, and z is the propagation axis of the photoassociation light. 
The observed Zeeman splitting is consistent with our expectations: the $J=2$ resonance is an E2 transition and $\Delta m = \pm 1$ is allowed for the external magnetic field along the x and the y axis, $\Delta m = \pm 2$ for the z axis.
In contrast, the $J=1$ resonance is an M1 transition and $\Delta m=\pm 1$ for the x and the y axis, $\Delta m =0$ for the z axis.

In summary, we report our successful observation of optical excitation into sub-radiant states of ${}^{174}\text{Yb}_2$ molecule.
The inherent weak excitation rate is overcome by tight-confinement in a three-dimensional optical lattice, which gains a bound-bound transition rate instead of free-bound.
The experimental measurements of binding energies, linewidth, and Zeeman shift confirm that the origin of the newly found resonances are the sub-radiant $1_g$ state.

\begin{acknowledgments}
We acknowledge Yb group members at Kyoto University for their experimental assistance.
This work was supported by the Grant-in-Aid for Scientific Research of JSPS (No. 18204035, 21102005C01 (Quantum Cybernetics), 21104513A03 (DYCE), 22684022), GCOE Program "The Next Generation of Physics, Spun from Universality and Emergence" from MEXT of Japan, FIRST, and Matsuo Foundation.
The research was partially supported by the Polish MNISW (Project No. N N202 1489 33), Nicolaus Copernicus University grant 141-F, and the Foundation for Polish Science TEAM Project cofinanced by the EU European Development Fund and is part of the program of the National Laboratory FAMO in Toru\'{n}, Poland.

\end{acknowledgments}

\bibliography{1g}
\end{document}